\documentclass[conference]{IEEEtran}
\usepackage{graphicx}
\usepackage{caption}
\usepackage{xspace}
\usepackage{amsmath}
\usepackage{amssymb}
\usepackage{xfrac}
\usepackage{subfigure}
\usepackage{balance}

\newcommand{\ieee}{\textsc{ieee}~802.15.4\xspace}

\newcommand{\imin}{$I_{min}$\xspace}
\newcommand{\imax}{$I_{max}$\xspace}
\newcommand{\I}{$I$\xspace}

\begin{document}

%\title{Topology Construction for \ieee Networks using Cross-Layer RPL}
\title{%Cross-Layer 
Topology Construction in RPL Networks over Beacon-Enabled 802.15.4}

\author{
	\IEEEauthorblockN{
		Mali\v{s}a Vu\v{c}ini\'{c}\IEEEauthorrefmark{1}\IEEEauthorrefmark{5},
		Gabriele Romaniello\IEEEauthorrefmark{1}\IEEEauthorrefmark{5},
		Laur\`ene Guelorget\IEEEauthorrefmark{1},
		Bernard Tourancheau\IEEEauthorrefmark{1},
		\\
		Franck Rousseau\IEEEauthorrefmark{1},
		Olivier Alphand\IEEEauthorrefmark{1},
		Andrzej Duda\IEEEauthorrefmark{1},
		and Laurent Damon\IEEEauthorrefmark{5}
%		\thanks{aaaa}
	}
	
	\IEEEauthorblockA{ 
		\IEEEauthorrefmark{1}Grenoble Alps University, CNRS Grenoble Informatics Laboratory UMR 5217, France
	}
	
	\IEEEauthorblockA{
		\IEEEauthorrefmark{5}STMicroelectronics, Crolles, France\\
		Email: \{firstname.lastname\}@imag.fr, \{firstname.lastname\}@st.com
	}
}

% There's nothing stopping you putting the seventh, eighth, etc.
% author on the opening page (as the 'third row') but we ask,
% for aesthetic reasons that you place these 'additional authors'
% in the \additional authors block, viz.
% \additionalauthors{Additional authors: John Smith (The Th{\o}rv{\"a}ld Group,
% email: {\texttt{jsmith@affiliation.org}}) and Julius P.~Kumquat
% (The Kumquat Consortium, email: {\texttt{jpkumquat@consortium.net}}).}
% \date{30 July 1999}
% Just remember to make sure that the TOTAL number of authors 
% is the number that will appear on the first page PLUS the
% number that will appear in the \additionalauthors section.

% correct bad hyphenation here
\hyphenation{sen-sors net-works low-power manu-facturers lite-rature}

\maketitle

\begin{abstract} 
  In this paper, we propose a new scheme that allows coupling \textit{beacon-enabled}
  IEEE 802.15.4 with the RPL routing protocol while keeping full compliance with
  both standards. 
  We provide a means for RPL to pass the routing information to Layer~2 before
  the 802.15.4 topology is created by encapsulating RPL DIO messages in beacon
  frames. 
  The scheme takes advantage of 802.15.4 command frames to solicit RPL DIO
  messages. 
  The effect of the command frames is to reset the Trickle timer that governs
  sending DIO messages. 
  We provide a detailed analysis of the overhead incurred by the proposed
  scheme to understand topology construction costs. 
  We have evaluated the scheme using Contiki and the instruction-level Cooja
  simulator and compared our results against the most common scheme used
  for dissemination of the upper-layer information in \textit{beacon-enabled}
  PANs.  
  The results show energy savings during the topology construction phase and in
  the steady state.
\end{abstract}

% % A category with the (minimum) three required fields
% \category{C.2.1}{Computer-Communication Networks}{Network Architecture and Design - Wireless Communication}
% %inspired from Theoleyre : http://clarinet.u-strasbg.fr/~theoleyre/uploads/Publis/pavkovic11rpl.pdf
% %A category including the fourth, optional field follows...
% % \category{D.2.8}{Software Engineering}{Metrics}[complexity measures, performance measures]

% \terms{Experimentation, Algorithms, Performance (TBC)} 

\noindent\begin{keywords}
IEEE 802.15.4, beacon-enabled mode, RPL, Wireless Sensor Networks, topology
construction, multi-hop networks.
\end{keywords}

\section{Introduction}
%Short intro to IP based sensor networks; discuss that higher level protocols have been standardized, citing Watteyne. State that RPL is becoming de-facto standard for routing. State that, on the other hand, industry using beacon-enabled 15.4 is having issues due to design choices of RPL (underlying mesh network) and inherent incompatibly with 15.4's cluster tree. State that most of the previous work focused on topology construction without taking into account the upper layer protocols and that our goal is to leverage benefits of available RPL implementations and to use its DAG for the creation of the cluster-tree. State contributions - without modifying any of the two standards (15.4 and RPL), we provide a way for RPL to run over a cluster tree; due to the Trickle algorithm, we significantly lower the control overhead exchanged in the network for L2 construction and as a consequence lower the energy consumption.
The long awaited Internet of Things (IoT) has never been closer. The industry has
fully begun to take part and the further development is all about standard
compliance. The upper layers of the IP protocol stack for Wireless Sensor
Networks (WSN) are being fine-shaped and the gaps between IETF and IEEE
standards \cite{154} are being bridged. Full solutions begin to emerge so product
interoperability and security are of primary concern \cite{oscar}. 

The \ieee standard~\cite{154} is widely recognized as the main
technology for low-power wireless sensor networking~\cite{watteyne}. Among its
different modes, we focus on the \textit{beacon-enabled} mode able to achieve
very low energy consumption by supporting a desired level of radio duty cycling
(the proportion between the periods nodes are on and off). In this paper, we
address the problem of running the Routing Protocol for Low Power and Lossy
Networks (RPL) \cite{rpl-rfc}, the IETF standard for routing in WSN, on top of
\ieee \textit{beacon-enabled} nodes.

The forwarding structure built by RPL is a Destination Oriented Directed Acyclic
Graph (DODAG). Each node keeps a list of available parent nodes closer to the
DODAG root (sink node) and selects one of them as the ``preferred parent'' based
on an objective function. When a link to the preferred parent fails, a node
switches to another parent in its list. 
At the link layer, the \textit{beacon-enabled} \ieee nodes need to construct a
cluster-tree anchored at the PAN coordinator (also the sink node) for supporting
multi-hop communication. 
Moreover, a node joining the cluster-tree has to associate with a coordinator (a
Layer 2 operation) before it may send any data frame. The choice of the
coordinator influences any possible choice of the RPL parent node. 
% As the design goal of RPL was to ensure the operation agnostic of
% Layer~2 (L2) technology, 
In the case of the \textit{beacon-enabled} \ieee nodes, the
problem is how to construct the 802.15.4 cluster-tree according to the RPL
routing information based on a DODAG. 

While both \textit{beacon-enabled} \ieee and RPL have been extensively studied within their
OSI abstraction layer, the joint operation is surprisingly still an open problem. 
The existing work in the literature \cite{bogdan-journal} requires extensive modifications 
to both standards, which is an
unrealistic requirement at the current stage of IoT stack development.

% Indeed, the problem has not passed unnoticed in the literature. While there is no doubt that the offered solutions bring performance benefits to RPL, they seem to overlook the importance of standard-compliance. The improvement in the behavior usually comes from heavy modifications of the underlying MAC protocol. From the practical standpoint, this is not acceptable. We discuss the related work in more detail in Section \ref{relatedwork}.

We propose a solution to the problem that satisfies the constraint of keeping
RPL and \ieee unchanged.
In our approach, RPL constructs its DODAG before the cluster-tree at L2 and
we use the RPL routing information (selection of the preferred parent) in the
association decision to establish links, i.e., to select the coordinator in the
cluster-tree that is the preferred parent in the DODAG. 
%Much work considered the topic of coordinator selection in the 802.15.4
%cluster-tree without considering the peculiarities of upper layers
%\cite{tdma-superframeschedule,muthukumaran09}. 

The proposed solution takes advantage of cross-layer signaling: a node joining
the network requests RPL information from neighbor 802.15.4 coordinators and
associates with the right coordinator based on the information in the RPL DIO
message (DODAG Information Object). 
We adapt the operation of the Trickle timer \cite{trickle-paper} that governs
the transmission of DIO messages to provide the required information to Layer 2
(the adaptation remains compliant with the RPL specification). 

The main contributions of the paper are the following:
\begin{itemize}
\item a new scheme that allows RPL to run over the \textit{beacon-enabled} \ieee without
  any modification to the two standards,
\item the scheme leading to energy savings both during the topology construction
  and in the steady-state, due to the use of the Trickle timer,
\item a simple probabilistic model of the Trickle timer and an analysis
  of the delay of the proposed scheme,
\item an evaluation of energy savings and the time for topology convergence
  based on the implementation of the proposed scheme in Contiki.
\end{itemize}

The paper is organized as follows. We present the background information on the
\textit{beacon-enabled} \ieee and RPL in Sections \ref{beaconed} and \ref{rpl}. We
provide a detailed description of the proposed scheme in Section \ref{proposal}
and evaluate it in Section~\ref{eval}. Section \ref{relatedwork} summarizes the
related work. We conclude and discuss the future work in Section
\ref{concl}.

\section{Beacon-enabled \ieee}
\label{beaconed}
%Briefly mention peer-to-peer mode of 15.4. State that beacon-enabled mode is a logical choice for networks where energy consumption needs to be guaranteed, i.e. harvested nodes. Explain the super frame structure; passive scanning, association procedure, active period scheduling for different clusters.
The lack of a radio duty cycling scheme in the \ieee \textit{non-beacon} mode has
led to intensive research on the \textit{beacon-enabled} mode. Periodic beacon
frames allow the synchronous sleep schedule of devices in the network. 
As a consequence, it is possible to \textit{guarantee} a desired level of
radio duty cycling (RDC), which is of the utmost importance for battery operated
nodes and especially for those that harvest energy from the environment. 
%BT lourd
%AD: I would delete this: we do not need to compare with other MACs and the
%comparison is complex - Simon says that ContikiMac should achieve the same RDC
%as beacon-enabled, but it suffers from costly broadcast. 
% Note that with typical RDC schemes based on preamble sampling that are often used in \textit{non-beacon} mode of \ieee, such as X-MAC, no guarantee exists and the RDC percentage is variable over time. 

The operation of nodes in the \textit{beacon-enabled} mode relies on beacons that
delimit the start of a {\em superframe}. 
Immediately following is the Contention Active Period (CAP) during which nodes
transmit pending data frames to their parent (cluster coordinator) using the
slotted CSMA/CA algorithm (a coordinator node needs to stay
active during CAP).
Beacon Order (BO) and Superframe Order (SO) are the key parameters to
tune the desired level of radio duty cycling in the \textit{beacon-enabled} mode.
Beacon Interval (BI) is defined as $BI~=~aBaseSuperFrameDuration~*~2
%AD: taken away "fully" - full duty cycle means 100%.
^ {BO}$ and $SD~=~aBaseSuperFrameDuration~*~2 ^ {SO}$ is the CAP duration. 
Thus, the upper-bound proportion of time a duty cycled coordinator node will
be active is $ 2^{SO-BO}$. 
Leaf nodes that only wake up for a transmission may benefit from an even lower
duty cycle. 

The network formed in the \textit{non-beacon} mode may be a mesh in which each node
may communicate with its radio-range neighbors, so running RPL in this case does
not raise any problems. 
Nodes in the \textit{beacon-enabled} mode have to form a \textit{cluster-tree}: a
node selects one parent node, the cluster coordinator, and synchronizes
with its beacons. 
The node may become a coordinator itself on behalf of other nodes, which enables
multi-hop communication from leaf nodes to the root of the cluster-tree. 
Scheduling of active periods of different coordinators is not defined in the \ieee standard
and is therefore left as an implementation choice.
%Much research concerned various strategies to avoid collisions of beacons from
%different coordinators (TDMA, FDMA, BOP etc.)
%\cite{tdma-superframeschedule,muthukumaran09,
%  multichannel-infocom,nazim-multichannel,BOP}.
In our evaluations, we use a simple static allocation of active periods that
allows us to focus on the topology construction without affecting the overall
results. 

%BT Fig.~\ref{superframe} illustrates an outgoing active period scheduled during the inactive part of the superframe.

%should look at 2 papers from nazim in the svn repository
% - 2011/multichannel-802_15_4/802154_MC.tex (cluster tree construction)
% - 2012/secon13-selfadapt/selfadapt-802154.tex (15.4 general presentation)

%\begin{figure}[ht!]
%
%\begin{center}
% \includegraphics[width=1\linewidth]{figures/superframe.pdf}
%	\caption{Outgoing and incoming superframes of a cluster coordinator}
%%	\vspace{-1.5em}
%	\label{superframe}
%	\end{center}
%\end{figure}
%\subsection{Topology construction phase}
%BT
%This section is of especial importance for understanding our results. In further text
%In the following, we assume that all the nodes in the network a-priori know the RF channel. In fact, the RF channel is often communicated to the nodes together with other information during network bootstrapping, using an out-of-band transmission, such as Near-Field Technology (NFC) or low-power RF transmission over all channels. 

The Personal Area Network (PAN) coordinator is the root of the tree, the
sink of the sensor network. 
It starts the topology construction by transmitting the first beacon. 
Other nodes are unassociated and have to switch their radio transceivers on to
perform passive scanning, the only mechanism for discovering potential
coordinators available in the \textit{beacon-enabled} mode. 
The reception of a beacon initiates a scan period during which a node waits for
beacons. 
At the end of this period, a node can initiate the association with the best
coordinator with the sequence of \texttt{association-request}, \texttt{ack},
\texttt{data-request}, \texttt{association-reply}, \texttt{ack} control frames. 

Note that most of the energy consumed during the topology construction
phase comes from idle listening during the scan period, which is unavoidable for
any association strategy that discovers the best available
coordinator. 
The duration of this interval should allow the discovery of all coordinators in
the radio range. 
%Recognizing that BO of 8 or less is suitable for most applications, in our studies we fix the duration of the ``scan period'' to 4 seconds \cite{bo7}.
%It is important to stress here that some approaches suggest complete avoidance of the scan period by associating to the originator of the first received beacon.
%%[MV]
%Such schemes do result in seemingly lower consumption during the topology construction phase. However, many authors in the literature have shown limitations of such a naive association policy once the network starts operating. Namely, resulting cluster-tree may be deep with many unnecessary hops. As a consequence, latency in the network will increase, while the effective throughput will be  lowered. 

\begin{figure}[htbp]
\begin{center}
 \includegraphics[width=0.95\linewidth]{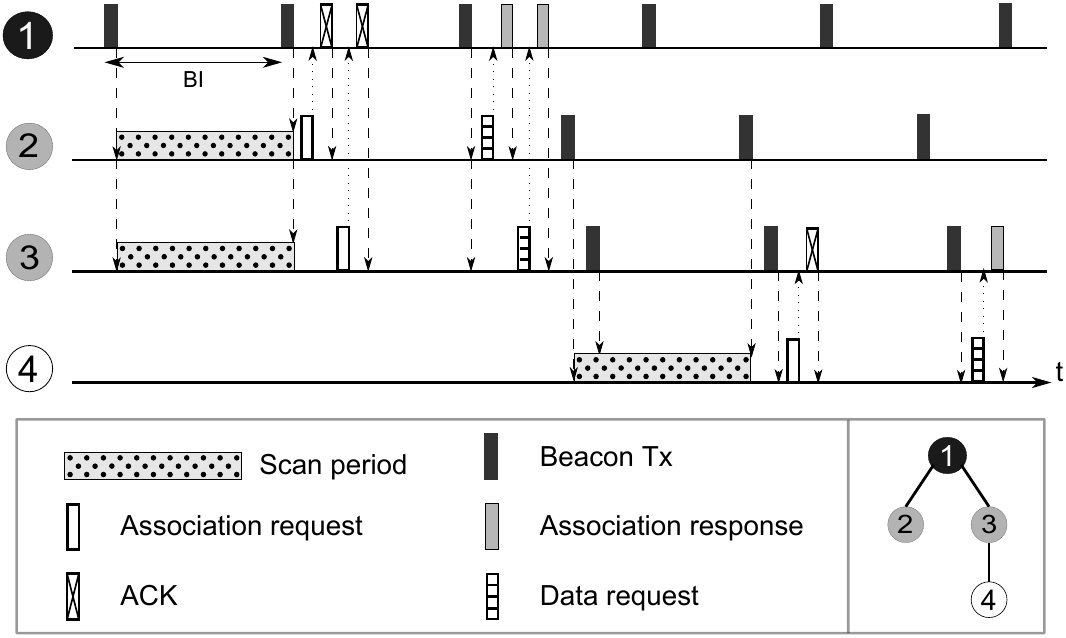}
	\caption{Topology construction in an example 802.15.4 cluster-tree.
\vspace{-0.6cm}}
	\label{assoc-procedure}
\end{center}
\end{figure}

%Note that if the ``scan period'' is skipped, and the node just associates with the first beacon it hears, some energy will be saved but the resulting topology might be far from optimal. 
Fig. \ref{assoc-procedure} illustrates a timeline of the topology construction
for an example cluster-tree composed of four nodes. 
Note that Node 4 may receive beacons from Coordinators 2 and 3, but it
selects Node 3 as the best parent. 

\section{RPL---Routing Protocol for Low Power and Lossy Networks}
\label{rpl}
%Level 3 routing protocol; Design choices of RPL, RPL messages. creation of DAG, objective function, preferred parent. 
Operating at the Network layer, RPL \cite{rpl-rfc} is a Distance Vector protocol that specifies how to construct a
Destination Oriented Directed Acyclic Graph (DODAG) with a defined objective
function and a set of metrics and constraints.

RPL specifies a set of new ICMPv6 control messages to exchange information related to a DODAG and notably:
\begin{itemize}
\item{\textit{DODAG Information Object (DIO)}} defines and maintains upward
  routes to the root, i.e. the DODAG. 
\item{\textit{DODAG Information Solicitation (DIS)}} messages
    request the DODAG related information from neighboring nodes.
%BT
%\item{\textit{DODAG Destination Advertisement Object (DAO)} advertizes prefix
%    reachability towards the leaf nodes of a DODAG enabling downward traffic.}
\end{itemize} 
The root starts the DODAG building process by transmitting a DIO. Neighboring
nodes process DIOs and make a decision on joining the DODAG based on the
objective function and/or local policy. A node computes its \textit{Rank} with
respect to the root and starts advertising DIO messages to its neighbors with
the updated information. As the process continues, each node in the network
receives one or more DIO messages and selects a preferred parent towards the
root. Note that for robustness, RPL keeps a list of other parents that can be
used in case link conditions change. 
As we focus on topology construction, we limit the discussion to upward routes. 
%AD: it's obvious and not needed for our problem.  
%Hence, RPL optimizes the upward routes for
%multipoint-to-point traffic that accounts for most of the traffic in LLNs.

%BT 
%To support downward routes, RPL uses DAO control messages that give the prefix
%information, the route lifetime, and other information about the distance of the
%prefix. 

%BT.
%As the design goal of RPL was to ensure the operation over an always connected
%Layer~2 (L2) technology, we
%need to find a means for a joint operation of RPL and \textit{beacon-enabled}
%802.15.4. 
In case of \textit{beacon-enabled} \ieee at L2, the traditional layer-independent operation
would confine the selection of RPL routes to those in the already-constructed L2 cluster-tree.
Consequently, the overall performance of RPL would be significantly degraded.
We exploit the approach of merging two structures: the 802.15.4 cluster-tree
and the DODAG of RPL, which allows us to benefit from low
overhead, small delays, and near optimal upward routes of RPL
\cite{rpl-performance-rfc} while creating the \ieee cluster-tree required for
low duty cycle communications. 
%Keeping RPL and 802.15.4 unchanged leads to the code reuse, which is of
%a particular importance in constrained environments. 

%\subsection*{The Trickle Algorithm}
%Explain how RPL uses the Trickle timer to schedule emission of DIOs. Brief the operation and parameters of the Trickle algorithm.
RPL uses the Trickle algorithm \cite{trickle-paper} to govern the transmission
interval of DIOs. The Trickle algorithm has three main parameters: i) the
minimum interval size \imin, ii) the maximum interval size \imax expressed as the
number of times the timer may double, iii) the redundancy constant $k$. 
%MV: I need Imax specifically described as the number of doublings for Section IVb.

The main idea of the algorithm is to exponentially reduce the amount of control traffic in the network when the topology is consistent, i.e. when there are no link failures or arriving nodes.
Consistency is checked by comparing the DODAG state advertised by other nodes in DIOs with the local one. If the number of consistent DIO receptions is higher
than redundancy constant $k$, Trickle refrains from transmitting.  
Instant $t$ at which Trickle decides if it is going to transmit is randomly
selected from interval [\I/2, \I), where $I\in~\{I_{min}~\times~2^{n}~\mid
n \in \mathbb~Z_{\ge 0}, n \le I_{max}\}$. Interval \I is doubled upon its
expiration by incrementing $n$. When a node detects inconsistency (which also
includes the initial DODAG construction), $n$ becomes $0$, which sets interval
\I to \imin. 
As detailed in the next section, we suitably adapt parameter \imin to construct
\ieee cluster-tree according to the DODAG of RPL. 

\section{
%BT Layer 2
802.15.4 Cluster-Tree Construction Based on RPL DODAG}
\label{proposal}
%Stress incompatibilities of RPL with 15.4 cluster-tree. Explain the passive scanning approach during which beacons from all neighboring coordinators are received. Explain the strategy that in order to communicate at L3 before being associated at L2, we put DIO messages as the beacon payload, This way, information is passed to the RPL sub-layer where the DODAG is independently built. Put a figure explaining how we put DIOs in the beacon payload, not affecting the Trickle timer (putting the DIO in the closest beacon). Explain the challenge of DIO size and that available beacon payload size is variable due to the number of downward messages (addresses of nodes communicated in the beacons). Make a simple calculation of worse case delay introduced to the Trickle timer this way. 

We propose the selection of the best coordinator in the 802.15.4 cluster-tree
based on the preferred parent in the DODAG of RPL. The resulting cluster-tree
will effectively be a subset of the DODAG initialized during the topology
construction phase. There are several issues with such an approach:
\begin{enumerate}
\item RPL is a network layer protocol, but no communication among nodes at Layer
  3 may take place before links at Layer 2 are established (node association
  with a coordinator).
\label{cross-point}
\item An  \ieee node once associated can only communicate with
  its cluster coordinator, so after association, a node can only receive DIO
  messages from its cluster coordinator.
\label{point2}
\end{enumerate}

In this paper, we focus on addressing the first issue to enable
topology construction. 
%AD: is it only the issue of having several coordinators at 802.15.4? For me
%it's a problem of having leaf nodes that are not coordinators and may have a
%better rank in DODAG. Is is a problem? I would say that in sucha configuration
%the cluster tree should be different and reflected at the DODAG. Just wonder if
%the second issue is a real problem...
The second issue is a
part of the future work that may add robustness to \ieee
cluster-tree by keeping alternate parents from the DODAG of RPL.

To address the first issue, we exploit the fact that DIO messages are
multicast. 
As Layer 3 multicasts translate to Layer 2 broadcasts, we use beacons to
broadcast DIO messages. 
There is no better broadcast mechanism in multi-hop \textit{beacon-enabled} networks than
the beacons themselves---during the scan period devices wait for beacons. 
We assume that \ieee Reduced Function Devices (RFD) are 
configured as RPL leaf nodes, i.e., they do not send DIO messages. 
Similarly, Full Function Devices (FFD) may become cluster coordinators, i.e.,
they have to be configured as RPL routers, which is a realistic assumption as
the role of a device mainly depends on its energy source. 
%AD: not good for "we keep standards unchanged" - why this require some change?
%A device knows whether it is RFD or FFD - if FFD, RPL is enabled.
%In case of harvested or battery powered nodes, the algorithm determining if a
%device should start beaconing may also pass this information to RPL, requiring
%minimal changes. 
We assume that a node a priori knows if it is an RFD or an FFD.
%BT

\begin{figure}[htbp]
\begin{center}
 \includegraphics[width=0.83\linewidth]{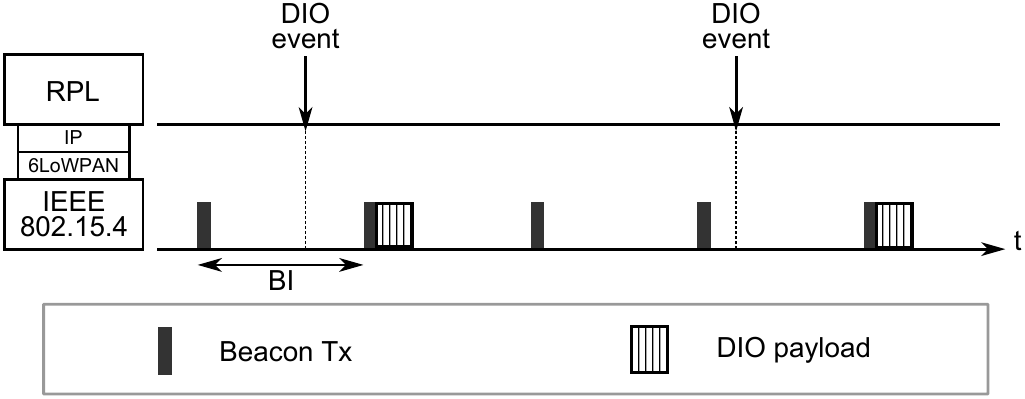}
	\caption{Encapsulation of DIO messages in beacon frames.\vspace{-0.3cm}}
	\label{beacon-payload}
\end{center}
\end{figure}

We propose the encapsulation of RPL DIO messages in the beacon frame payload
following an idea discussed in the team \cite{franck-report}. 
%AD: already said.
% As our goal is to keep the logarithmic scalability and low overhead of Trickle,
% we leave the RPL layer unchanged. 
Layer 2 adds DIO to the payload of the next scheduled beacon if the resulting
frame does not exceed \ieee maximum physical layer frame size of 127 bytes
(cf. Fig. \ref{beacon-payload}). 
In case the DIO message cannot fit into the current beacon, it may be fragmented
or delayed for the following one as the beacon payload size varies as a function of downward
traffic. 
%AD: independent, but requires modification of Layer 2.
% Note that this approach, illustrated in
% Fig. \ref{beacon-payload}, keeps the two network layers' operation independent.

%AD: replace "DIO arrival" by "DIO transmission" - it is the instant at which
%RPL decides to send a DIO. 
%

The exponential increase of the DIO transmission interval governed by Trickle
has an important side effect: arriving nodes would potentially wait a long
time interval before receiving the first DIO. 
RPL addresses this issue with DIS messages that can be broadcast to solicit
the transmission of a DIO: upon
reception of a DIS, a node  resets its Trickle interval \I to
\imin so DIO will be transmitted shortly \cite{rpl-rfc}. 
%AD: not needed for explanation.
%A DIS message is typically transmitted at node boot time. 
However, DIS broadcast is not enough for synchronous duty cycled
networks---neighbor nodes in the radio range may sleep at the
instant of the DIS transmission. 
As explained above, the reception of a beacon delimits the start of the
Contention Active Period during which the coordinator is active. 
Thus, CAP is the most suitable period during which an unassociated node may
solicit information from nearby coordinators. 
Note that a node wanting to join the network is awake during the scanning period so it can
receive beacons from several neighbor coordinators. 
Thus, we propose that the node
transmits a solicitation message by performing CSMA/CA after the beacon if the
following two conditions hold:
\begin{itemize}
\item the received beacon is the first one received from a given coordinator, 
\item the beacon does not contain a DIO in its payload.
\end{itemize}

The solicitation message could be a RPL DIS message encapsulated in an
802.15.4 command frame. 
Note that a node cannot send data frames before association \cite{154}.
However, we have chosen to use the \ieee \textit{beacon-request} command frame 
without any payload as a solicitation message---it has a small size (8 bytes) so a very short
transmission time. Additionally, the RPL specification \cite{rpl-rfc} allows the Trickle reset triggered by external events.
%AD: this is not an argument for chosing beacon-request.

%MV: it kind of is because of difference in size (8 bytes vs. 64 bytes). plus, DIS 
% can be encapsulated in the payload of any command frame, including beacon request.
% see the sentence added for clarification.

% In fact, current draw for RX and TX modes of modern radio
% chips is approximately the same. Thus this makes no difference in terms of
% energy consumption if a device is transmitting or receiving. However, during a
% transmission the device is deaf such that it is beneficial to minimize the time
% spent in TX mode.
%BT.
Note that the \textit{beacon-request} command frame is typically used in the
\textit{non-beacon} mode to solicit the information about the network. 
It has no use in the \textit{beacon-enabled} mode as beacons are periodically transmitted. 
We use its reception at Layer 2 to trigger the reset of the Trickle timer at the
RPL layer to spawn a DIO transmission. 
The goal is to encapsulate the DIO message in the following beacon so that
arriving nodes can select the best coordinator. 
As a node may send several \textit{beacon-request} solicitation frames during
the scan period (and CAP of each detected coordinator), the scheme ensures the
reset of the Trickle timer for all RPL routers in the range. 
 
A possible drawback of the scheme could be its possible side effect on the
duration of the always-on scan period. 
In fact, with typical parent selection schemes at Layer 2, 
% [MV] To add a reference.
each beacon carries a network-specific metric processed by arriving nodes.
Then, in case Beacon Order is a priori known, the worst-case scan duration is
one Beacon Interval. 
However, a simple algorithm achieves the same duration with our scheme
as well---during the scan period of duration $BI$:
 \begin{enumerate}
 \item for each discovered coordinator, a node stores the expected instant of
   the next beacon ($current\_time()~+~BI$),
 \item for each discovered coordinator, a node solicits the reset of the Trickle
   timer as explained above,
 \item upon expiration of $BI$, a node goes to sleep and schedules its wake up
   at the instants found in (1),
 \item a node wakes up and receives the beacon with the DIO payload,
 \item upon reception of the DIO payload from the last discovered coordinator, a
   node consults RPL about the best choice and schedules the next wake up just
   before the beacon of the selected coordinator; then, the node follows
   the standard association procedure.
 \end{enumerate}
 
\begin{figure}[ht!]
\begin{center}
 \includegraphics[width=0.95\linewidth]{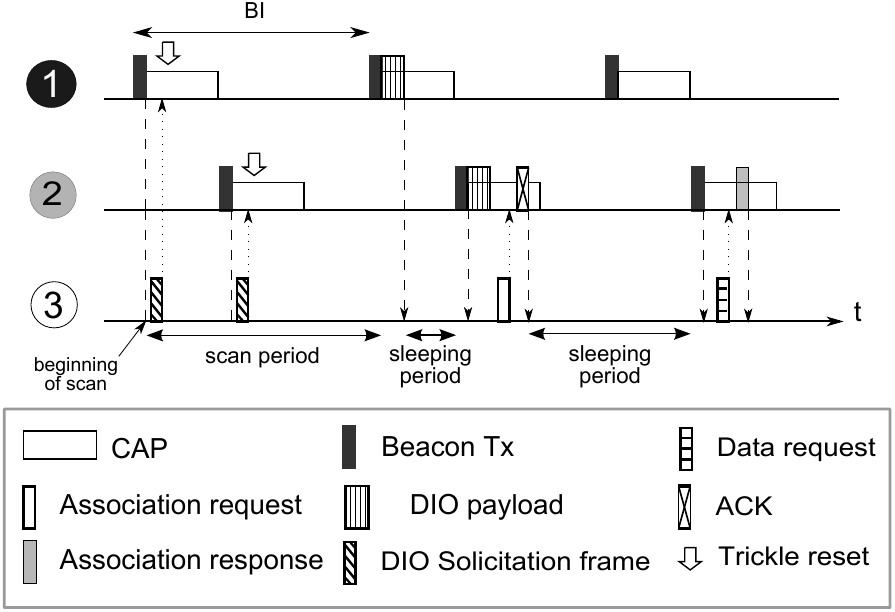}
	\caption{Soliciting DIO during the scan period. 
	%RFD node (3) is in the radio range of both (1) and (2) but selects (2) as the best coordinator.
	%BT already in the text
	\vspace{-0.6cm}}
	\label{beaconrequest-solicitation}
\end{center}
\end{figure}

%BT In fact, the scheme will ensure
This scheme ensures the discovery of all coordinators in the radio range while
allowing a node to start duty cycling after one $BI$ from the boot time
(cf. Fig. \ref{beaconrequest-solicitation}). 
During next $BI$, node receives DIOs and passes them to RPL. 
In the worst case, by the end of the second $BI$, RPL will have the preferred
parent selected.  
The additional \textit{worst-case} delay of one $BI$ is the price to pay during
the topology construction for the benefit that comes later-on with the Trickle
timer during the network operation. 
As the node spends most of the second beacon interval sleeping, it consumes energy
only for receiving beacons. 
For $n$ discovered coordinators, the energy will be $E = n \times T \times
I_{RX} \times V$, where $ I_{RX}$ is the radio current draw in receive mode, $V$
the operating voltage, and $T$ transmission time of one \ieee beacon with a DIO
message in its payload (typically around 3.5 ms for 250 kb/s \ieee compliant radios).

%BT
%AD: BO in beacons. 802.15.4 sets the same BO in the whole network. Thus, why
%different values of BO?
%MV: What if a node does not know the PAN it wants to associate and/or RF channel...
Note, however, that in many deployments, $BO$ is not a-priori known.
%MV: In some Zigbee spec Gabriele passed me, they discuss that most applications
% require BO from 4 to 8. Otherwise, passive scan over all channels becomes completely unfeasible...
%AD: 4 seconds is not the largest expected BI. Should be Max (BIi), no?
In such cases, devices have to scan for longer periods to account 
for the largest expected $BI$ in presence of multiple PANs \cite{karowski}. 
Our scheme in such scenarios introduces no additional delay as long as the
preconfigured scan duration is greater than or equal to half the actual $BI$ in
the network. 
%AD: already said.
%Fig. \ref{beaconrequest-solicitation} illustrates the proposed scheme.

\subsection{\imin Parameter Tuning and Analysis}
\label{analysis} 

The successful operation of the proposed scheme requires that, upon
solicitation, the subsequent beacon includes a DIO message. 
To achieve such behavior while keeping the operation of two layers independent,
we need to configure the Trickle \imin parameter as a function of $BI$, because
the reception of a solicitation frame triggers the Trickle timer reset and  
the next timer value will be uniformly drawn from the interval
[\imin/2,~\imin). 
Thus, to ensure the arrival of the next DIO before the
subsequent beacon, the following condition needs to hold:
\begin{equation}
\label{firstcondition}
I_{min} \le BI - SD,
\end{equation}
where $SD$ denotes CAP duration. 
%AD: notation clash - SD is CAP duration...
Similarly, as previously discussed, the worst case scan period
when $BO$ is a priori known, is $BI$. 
The optimal performance of Trickle with our scheme is obtained when $I_{min} =
BI - SD$, which ensures the successful operation while having the lowest overhead. 

%BT
\subsection{Analysis of DIO Reception Delay}

\begin{figure}[htbp]
\begin{center}
 \includegraphics[width=0.75\linewidth]{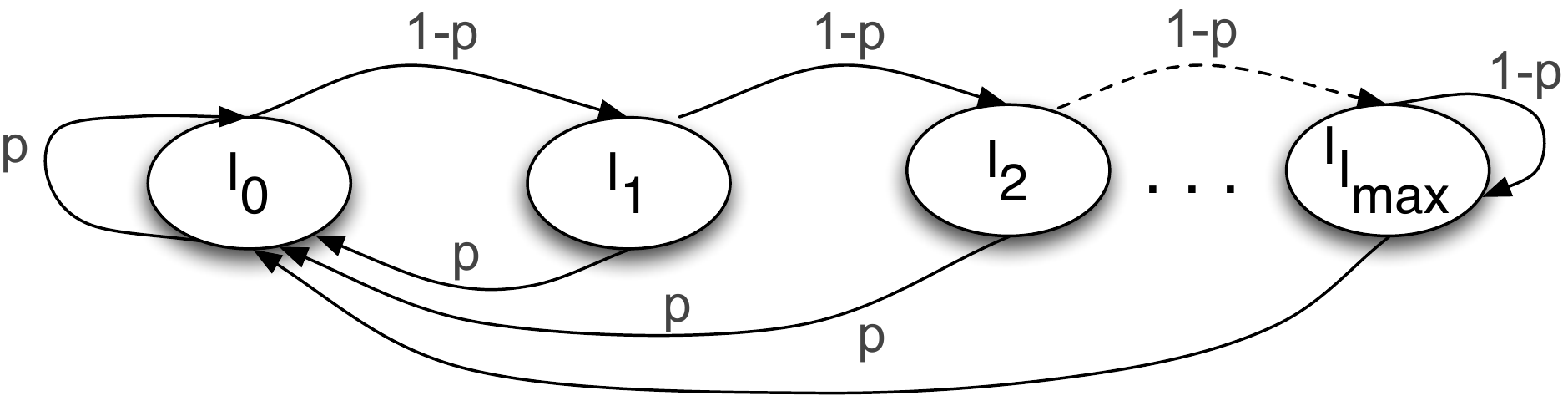}
	\caption{Markov chain with $I_{max}$+1 states for Trickle. 
	%BT interval $I$.
	\vspace{-0.6cm}}
	\label{markov-chain}
\end{center}
\end{figure}

We evaluate here the expected delay of DIO messages encapsulated in periodic
beacons. We define the Trickle timer value as random variable $X$ uniformly
distributed in [\I/2, \I), where $I$ is a random variable denoting the current
Trickle state. Then, from the Layer 2 point of view, a DIO message arrives
during a beacon interval at instant $X~mod~BI$. Delay $D$ is the interval
remaining until the transmission of the next beacon:
%\begin{align*}
%D = BI - (X~mod~BI).
%\end{align*}
%Using Knuth's floored division this can be written as:
\begin{equation}
D = BI - (X - \left\lfloor\frac{X}{BI}\right\rfloor *BI).
\end{equation}
The expected delay is then:
\begin{equation}
\label{favorite}
E[D] = BI - E[X] + E[\left\lfloor\frac{X}{BI}\right\rfloor] * BI.
\end{equation}
Now, recall that $I$ is a discrete random variable in
$\{I_{min}~\times~2^{n}\}$, where $n = 0, 1, \dotsc, I_{max}$.  
% [MV] Note sure if this with Markov chain is needed since we only derive for the special case n = 0;
We model $I$ with a discrete-time Markov chain shown in Fig. \ref{markov-chain},
where $p$ denotes the probability of the Trickle reset.
%BT
We can notice from Fig. \ref{markov-chain} that stationary probabilities of states $I_{0}, \dotsc,  I_{I_{max}-1}$ follow a geometric distribution with reset probability $p$:
%\begin{align*}
$\Pi_{I_i}~=~(1~-~p)^ip, i = 0,\dotsc, I_{max}-1.$
%\end{align*}
%BT
The last state, $I_{I_{max}}$ has the stationary probability:
%\begin{align*}
$\Pi_{I_{I_{max}}} = (1-p)^{I_{max}}.$
%\end{align*}
We can find the expected Trickle timer value as $E[X] = E[E[X | I]]$.

As our scheme uses the \textit{beacon-request} solicitation frame at L2 to reset
Trickle, the case $I = I_{min}$ is of a particular interest. 
From Eq. \ref{favorite}, it follows that:
\begin{equation}
E[D | _{I = I_{min}}] = BI - E[X |  _{I = I_{min}}] \\+ E[\left\lfloor\frac{X}{BI}\right\rfloor | _{I = I_{min}}] * BI.
\label{genericeq}
\end{equation}
Given the condition of Eq. \ref{firstcondition} and also the fact that the right endpoint is excluded from the uniform interval, term $E[\lfloor\frac{X}{BI}\rfloor]$ goes to zero leaving:
%\begin{align*}
$E[D~|~I~=~I_{min}]~=~BI~-~E[X~|~I~=~I_{min}].$
%\end{align*}
Finally, as $X$ is now a uniform random variable in $[I_{min}/2, I_{min})$, the
expected DIO delay becomes:
\begin{equation}
\label{finaleq}
E[D | I = I_{min}]  = BI - \frac{3}{4}I_{min}, ~~I_{min} \le BI.
\end{equation}

We have validated Eq. \ref{finaleq} by emulating a real node running the Contiki
operating
system  for constrained devices.
We have timestamped the expiration instants of Trickle and the instants of the
beacon with DIO transmission. 
We have configured $I_{min}$ to an approximate value of $BI/2$ (Contiki accepts the
values of $I_{min}$ in power of 2). 
The emulation results over 5000 samples strongly corroborate our analysis
with a maximal error of 2.799\%. 
% MV CHECK EXACTLY THE ERROR

%on 99% confidence intervals values.%Fig.~\ref{diodelay} presents the obtained results. Each measured point is averaged over 5000 values and presented with 99\% confidence intervals.

% \begin{figure}[ht!]
%\begin{center}
% \includegraphics[width=0.95\linewidth]{figures/diodelays_fbo.pdf}
%	\caption{Expected and measured delay introduced to Trickle for $I = I_{min}$.}
%	\label{diodelay}
%\end{center}
%\end{figure}

From Eqs. \ref{firstcondition} and \ref{finaleq}, it follows that for setting $I_{min} = BI - SD$, our scheme introduces
the least additional delay to Trickle after reset, while ensuring successful operation.
%Explain alternative approach with OPT messages.

 \begin{figure}[htbp]
\begin{center}
 \includegraphics[width=0.60\linewidth]{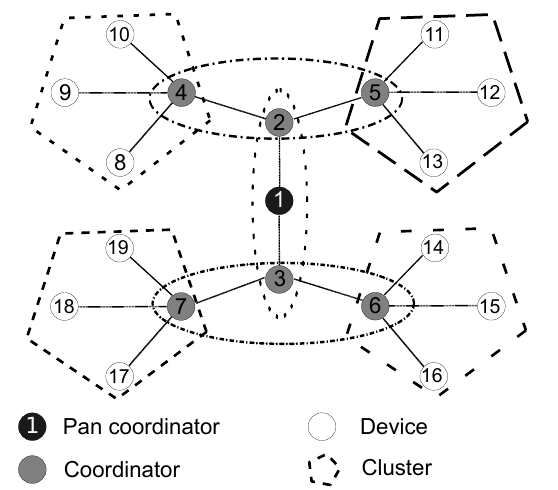}
	\caption{Evaluated topology.\vspace{-0.6cm}}
	\label{topology}
\end{center}
\end{figure}

\section{Performance Evaluation}
\label{eval}
%Discuss the simulation topology, instruction-level emulations in Cooja; present results and discuss improvements; state that for delay we were aiming to achieve comparable delays as with OPT messages. (need to find appropriate name for OPT messages though...)
%We evaluated our proposal using operating system Contiki and the instruction-level emulator Cooja. Note that the only imperfections of Cooja in respect to the real world environment come from the radio channel model - Unit Disk Graph. 

%[MV] Some parts taken from Multichannel hello paper. Is it ok?!?
To evaluate our scheme, we have used an implementation of the \ieee \textit{beacon-enabled} mode specifically developed and optimized 
for harvested sensor motes manufactured by STMicroelectronics (ST) containing a 32 bit microcontroller and a proprietary radio 
802.15.4 transceiver. To our knowledge, it is the first \ieee \textit{beacon-enabled} implementation for the Contiki operating system.
To benefit from the Cooja simulator \cite{cooja-paper} that uses the MSPsim instruction-level emulator of the Tmote Sky platform, our team ported the \textit{beacon-enabled} layer developed for the ST motes to the Tmote Sky platform.
%BT ICI mettre un \thanks ?
Tmote Sky is based on a 16-bit  MSP430 microcontroller, operating at 8MHz clock rate, and the CC2420 Chipcon radio. Note that the only imperfection of Cooja with respect to the real world environment comes from the Unit Disk Graph radio channel model. Fig. \ref{topology} presents the evaluated topology.

 %We implemented and compared the proposed scheme against a Layer 2 topology construction mechanism that selects the preferred parent based on the hop distance from the sink - named in our team as "OPT".  Information about hop distance is similarly encapsulated in the beacons, without any scheduling - i.e. each beacon carries an OPT message to allow the fastest coordinator selection (as each beacon carries an OPT message, algorithm selects the closest coordinator to the sink immediately after the scan period). Note that the OPT messages are \textbf{2.32 times} smaller than the corresponding DIO. Format and size of DIO messages is specified by the RPL specification \cite{rpl-rfc}, while OPT messages contain only the most necessary information and are not encumbered by the standard. 
%AD: need to say that leaf nodes are RFD and others are FFD.

\begin{figure}[htbp]
\noindent
\centering
\subfigure[Overhead, variable SBP size.]{
\includegraphics[width=0.48\columnwidth]{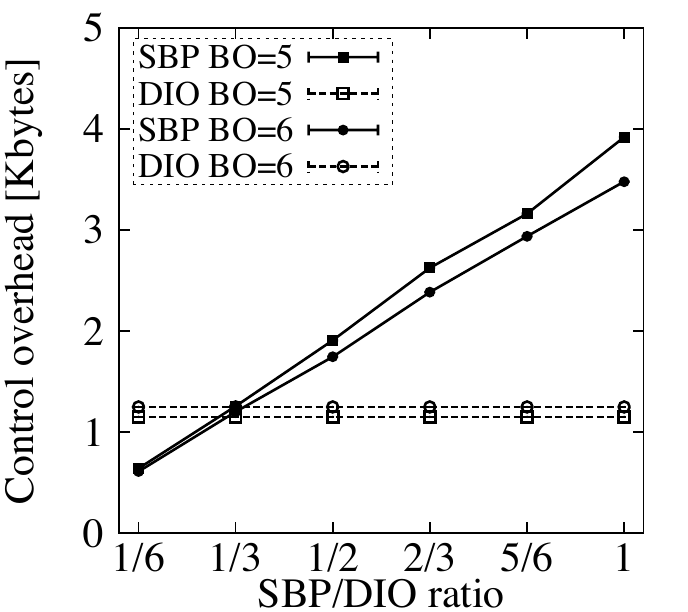}
\label{overhead-ratio}
}
\hspace{-0.37cm}
\subfigure[Topology convergence time.]{
\includegraphics[width=0.48\columnwidth]{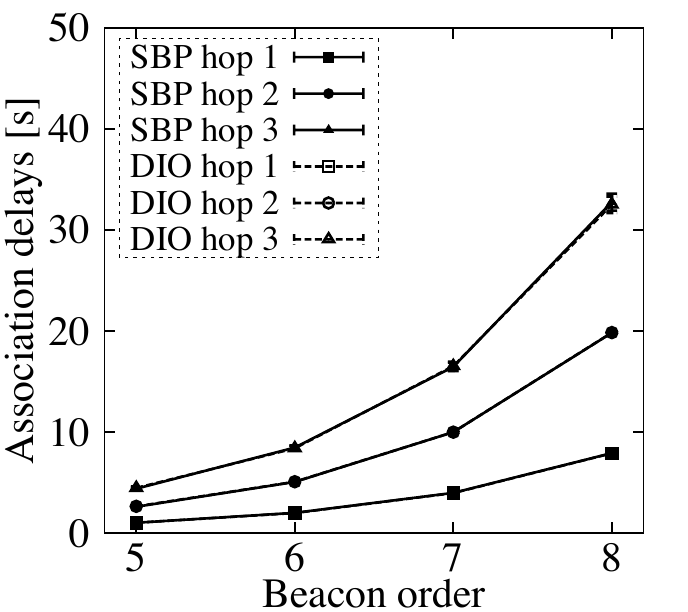}
\label{delay-case1}
}
% \begin{figure}[ht!]
%\begin{center}
% \includegraphics[width=0.8\linewidth]{figures/delays_fbo_phase1.pdf}
%	\caption{Topology convergence time.}
%	\label{delay-case1}
%\end{center}
%\end{figure}

% \begin{figure}[ht!]
%\begin{center}
% \includegraphics[width=0.8\linewidth]{figures/overhead_fbo_phase1.pdf}
%	\caption{Total control overhead during topology construction when SBP = 1/3 DIO.}
%	\label{overhead-case1}
%\end{center}
%\end{figure}
%
% \begin{figure}[ht!]
%\begin{center}
% \includegraphics[width=0.8\linewidth]{figures/energytx_fbo_phase1.pdf}
%	\caption{Energy spent in TX during topology construction when SBP = 1/3 DIO.}
%	\label{energytx-case1}
%\end{center}
%\end{figure}
%
% \begin{figure}[ht!]
%\begin{center}
% \includegraphics[width=0.8\linewidth]{figures/energyrx_fbo_phase1.pdf}
%	\caption{Energy spent in RX  during topology construction.}
%	\label{energyrx-case1}
%\end{center}
%\end{figure}
%I had to remove RX/TX abbreviations in order for it to fit...
\subfigure[Energy spent in transmission, in mJ.]{
\includegraphics[width=0.48\columnwidth]{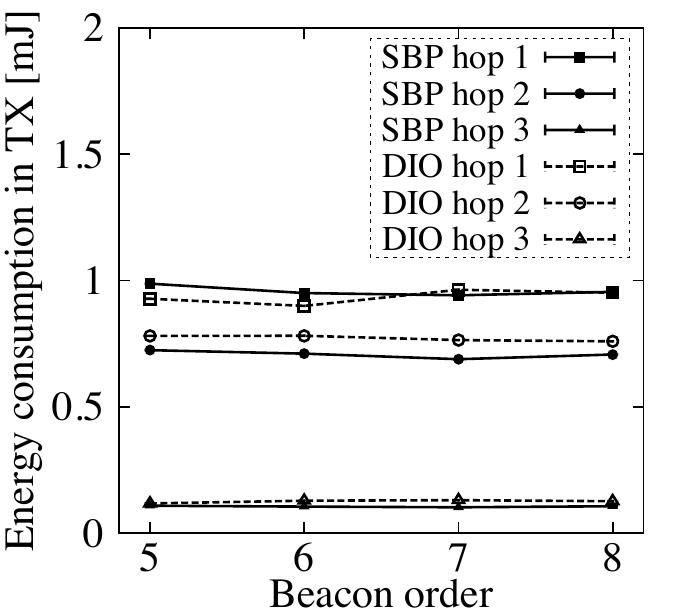}
\label{energytx-case1}
}
\hspace{-0.37cm}
%I had to remove RX/TX abbreviations in order for it to fit...
\subfigure[Energy spent in reception, in J.]{
\includegraphics[width=0.48\columnwidth]{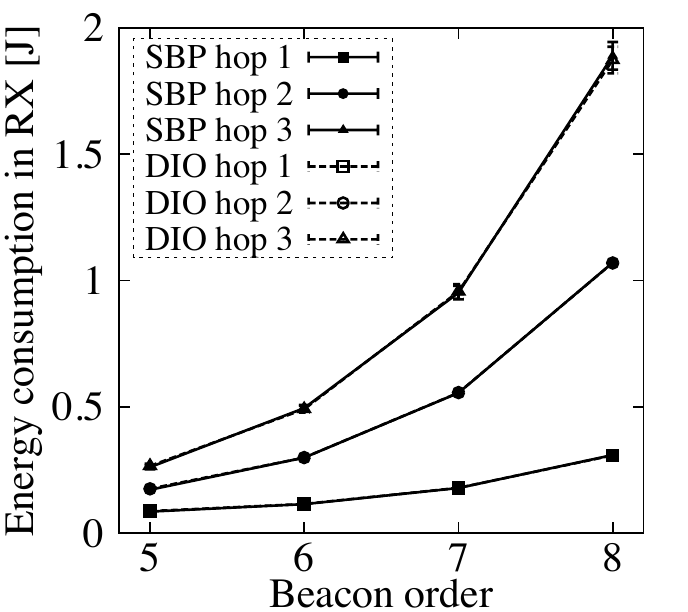}
\label{energyrx-case1}
}
\vspace{-0.2cm}
\caption{Results from emulation during topology construction.\vspace{-0.2cm}}
\end{figure}

Many authors in the literature discussed the method of encapsulating information necessary
for topology construction in the beacon payload (parent selection, neighbor discovery)
\cite{kohvakka,bogdan, zigbee-beaconpayload}. Consequently, they assume the information 
to be present in each beacon. As our goal in this paper was to present benefits in terms of 
802.15.4 topology construction, we have compared our scheme against this approach
and denote the scheme Systematic Beacon Payload (SBP). 
% [MV] add references.
To be fair and not to loose the generality of our results, we have
studied the effects of varying the SBP message size and how it affects
performance. We found that the two schemes have similar performance when the SBP
message size is approximately 1/3 of the DIO size
(cf. Fig.~\ref{overhead-ratio}), that is, when one coordinator from
Fig. \ref{topology} sends 1 DIO message for every 3 beacons with SBP on the
average during topology construction. Note that this ratio depends on the
duration of the scan period and the configuration of Trickle. 
%Therefore, it should not be generalized.
%BT
For a given implementation, one can easily evaluate such a ratio and derive the
gain or loss depending on the message size parameters. 

%MV: Due to imperfect CSMA/CA I had to account for this before, so the code had a 
% condition to effectively reduce Imin as a function of BI, in order to guarantee the arrival
% of DIO in the next beacon.

We set the $I_{min}$ Trickle parameter to approximately $BI~-~SD$ and keep SO
equal to 2. We compute the radio
energy consumption from the current draw values reported in the Tmote Sky data
sheet.  
We average all the points in the following graphs over 20 emulation runs and
show them with 95\% confidence intervals.

%For case (1), we found the two schemes to perform equally for SBP message size that is 1/3 of DIO size. That is, for topology in Fig. \ref{topology} on average there is 1 DIO message every 3 beacons with SBP -- during topology construction. However, this is not true for case (2) as due to the sub-optimal scan duration, there are more beacons unnecessarily sent before the nodes associate.

We can notice in Fig. \ref{topology} that nodes have only one coordinator in
their radio range. 
We have chosen such topology to focus on topology construction in RPL networks
over \textit{beacon-enabled} 802.15.4 and evaluate the effect of our scheme.
%BT is this the worst case?
%MV in terms of solicitation yes, because if a solicitation fails following beacon does not have a DIO and the node has to start a new scan period. In terms of delay it is best-case because we do not introduce any delay more than SBP. It is explained bellow. Let me know if it needs additional clarification...
In this way, we isolate topology construction aspects from the problems related
to routing that may depend on the choice of routing metrics or objective
functions. 
Moreover, a single coordinator discovered during the scan period $BI$ means that
the solicitation scheme is put under stress.
Indeed, if a single DIO message does not arrive with the subsequent beacon upon
solicitation, the node will have to initiate another scan period, which would
unnecessarily increase the topology convergence delay. 
%BT can you explain ?
%MV should be ok now?
Nevertheless, the example topology in Fig.~\ref{topology} is % somehow
%favorable to the proposed scheme---it does not introduce an additional delay in
favorable to the proposed scheme in terms of delay---it does not introduce additional delay in
case $BO$ is a priori known, i.e., the first discovered coordinator is also the
last one, so a node can initiate the association procedure after the scan period
of one $BI$. 
However, we discuss the worst case delay in the presence of multiple
coordinators in Section \ref{proposal}. 

\vspace{-0.2cm}

\vspace{0.2cm}
Also note that in some cases, the first beacon discovered during the scan
period may already contain a DIO message. 
As the Trickle timer randomly selects its expiration interval and our scheme
keeps the operation of two layers independent, it is a lucky outcome. In this
case, a node does not need to solicit DIO as detailed
in Section \ref{proposal}. However, a node still has to wait for the expiration
of the scan period before initiating its association procedure to ensure that it
has discovered all potential coordinators.

\vspace{-0.2cm}
\begin{figure}[htp]
\noindent
\centering
\subfigure[Total overhead]{
\includegraphics[width=0.48\columnwidth]{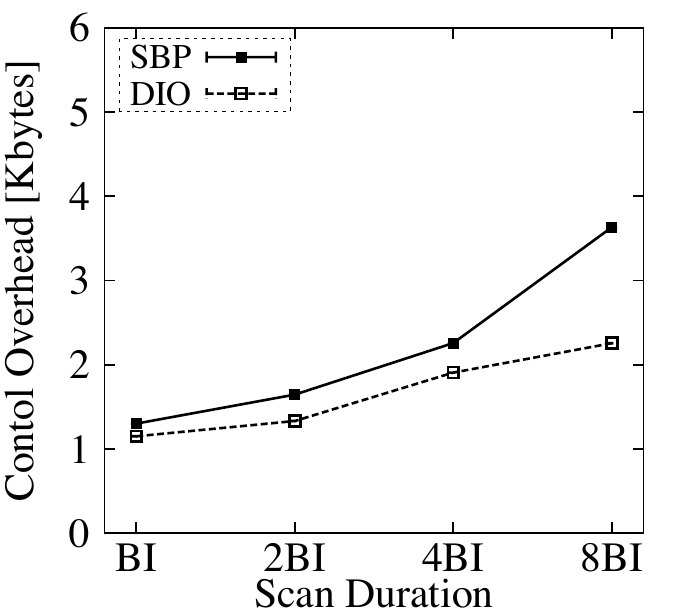}
\label{overhead-fsd-case1}
}
\hspace{-0.37cm}
% \begin{figure}[ht!]
%\begin{center}
% \includegraphics[width=0.8\linewidth]{figures/overhead_fsd_bo5.pdf}
%	\caption{Total control overhead during topology construction for variable scan duration and BO 5.}
%	\label{overhead-fsd-case1}
%\end{center}
%\end{figure}
\subfigure[Energy spent in transmission, in mJ]{
\includegraphics[width=0.48\columnwidth]{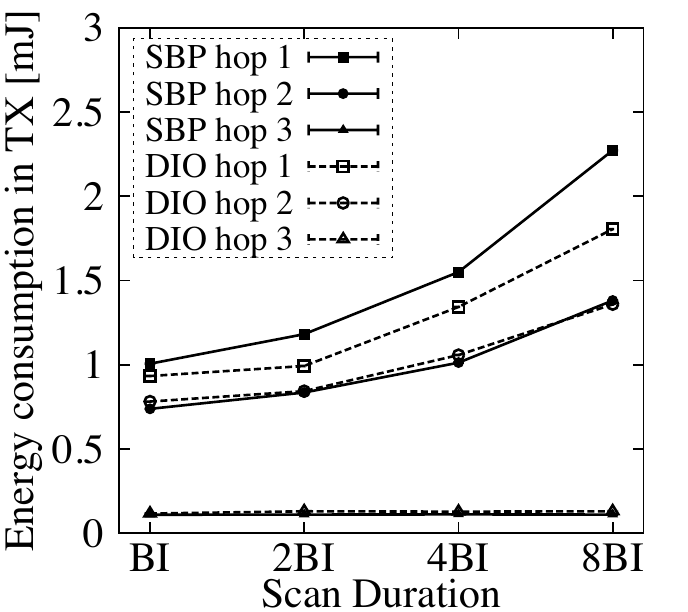}
\label{energytx-fsd-case1}
}
% \begin{figure}[ht!]
%\begin{center}
% \includegraphics[width=0.8\linewidth]{figures/energytx_fsd_bo5.pdf}
%	\caption{Energy spent in TX during topology construction for variable scan duration and BO 5.}
%	\label{energytx-fsd-case1}
%\end{center}
%\end{figure}
\vspace{-0.2cm}
\caption{Results from emulation during topology construction for variable scan duration and $BO= 5$.\vspace{-0.2cm}}
\end{figure} 

We present the results for the case in which two schemes have the most similar
performance, i.e., we set the message size of SBP to 1/3 of 
DIO (cf.  Fig. \ref{overhead-ratio}). 
%BT (respectively smaller) (respectively better).
Larger SBP message sizes result in worse performance while smaller SBP
messages result in better performance during topology construction in case
$BO$ is a priori known. 
 
\vspace{-0.1cm}
\subsection{Topology Construction}\vspace{-0.1cm}

We study the topology construction phase for two cases: 1) $BO$ is a priori
known so the scan period can be set to the minimal value of $BI$;
2) there is no a priori knowledge of $BO$ so nodes use a sub-optimal scan
duration to account for the worst case. In both cases, simulations last until
the association of the last node. 

For case 1), Figs. \ref{delay-case1}-\ref{energyrx-case1} present the
results for varying $BO$. 
We can see in Fig. \ref{delay-case1} that our scheme does not introduce any
additional delay for the evaluated topology and the results for two schemes are
similar within confidence intervals. 
Fig. \ref{energytx-case1} shows similar results in terms of cumulative energy spent in transmission, a
consequence of the choice of the parameters for two schemes. 
Notably, coordinators at hop 1 and 2 spend approximately the same energy transmitting beacons. 
The major part of the energy spent in reception comes from idle listening during
the scan period so two schemes perform equally (cf. Fig.~\ref{energyrx-case1}).

For case 2), when $BO$ is not a priori known, we vary the scan period. 
As nodes remain in reception mode much longer, the energy spent in reception
makes the major part of the total consumption. 
Similarly to Figs. \ref{delay-case1} and \ref{energyrx-case1}, two schemes
perform equally. 
However, as the scan period is longer, there is a larger number of beacons
transmitted before the topology converges. We can thus see the effect of
the Trickle algorithm and the proposed solicitation scheme
(cf. Fig. \ref{overhead-fsd-case1}) that results in energy savings for hop 1
nodes as they transmit beacons the longest until the end of the tree
construction (cf. Fig. \ref{energytx-fsd-case1}).

\vspace{-0.1cm}
\subsection{Steady-state}\vspace{-0.1cm}

\begin{figure}[htbp]
\noindent
\centering
\subfigure[Energy spent in transmission, in mJ.]{
\includegraphics[width=0.48\columnwidth]{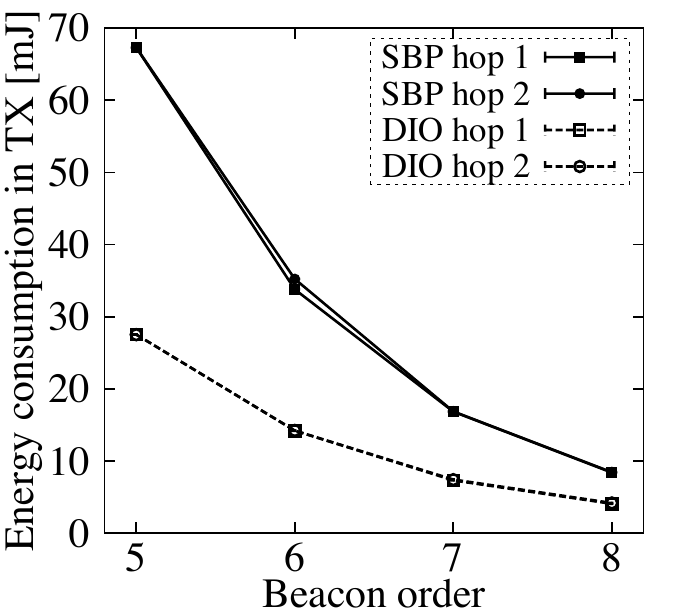}
\label{energytx-steadystate}
}
\hspace{-0.37cm}
% \begin{figure}[ht!]
%\begin{center}
% \includegraphics[width=0.8\linewidth]{figures/energytx_steadystate.pdf}
%	\caption{Energy spent in TX during 6 minutes of network operation in the steady state.}
%	\label{energytx-steadystate}
%\end{center}
%\end{figure}
%
%I had to remove TX/RX abbreviations in order for it to fit...
\subfigure[Energy spent in reception, in J.]{
\includegraphics[width=0.48\columnwidth]{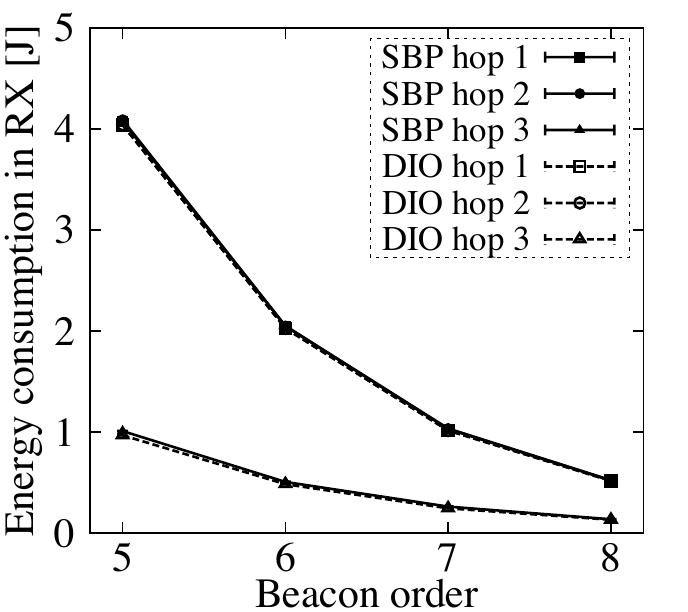}
\label{energyrx-steadystate}
}
% \begin{figure}[ht!]
%\begin{center}
% \includegraphics[width=0.8\linewidth]{figures/energyrx_steadystate.pdf}
%	\caption{Energy spent in RX during 6 minutes of network operation in the steady state.}
%	\label{energyrx-steadystate}
%\end{center}
%\end{figure}
\vspace{-0.2cm}
\caption{Results from  emulation during 6 min. 
%BT of network operation in the 
of steady state.\vspace{-0.5cm}}
\end{figure}

Furthermore, we have evaluated the benefits in terms of energy savings in the
steady state, i.e., after topology construction. 
There was no application traffic in the network and nodes simply 
duty cycle according to their schedules. 
The presented results concern 6 minutes of the network operation after the
association of the last node. 
We can see the effect of the reduction in control overhead by the Trickle
algorithm in Fig. \ref{energytx-steadystate}. 
In particular, FFD nodes (hop 1 and 2) transmit short beacons without any
payload most of the time, which results in energy savings both during reception
and transmission. 
%AD: which effect? It's not clear...
%MV: Factor of 2.3  (BO 5) reduction in TX energy spent in Fig 8a...
During reception, however, a major part of energy consumption comes from active
listening during the CAP of each coordinator so this effect is masked
(cf. Fig. \ref{energyrx-steadystate}). 
Note that in Fig. \ref{energytx-steadystate}, the consumption of RFDs is
zero as there is no application traffic  in the network. 
Also, during the steady state, the reception consumption of FFDs (hop 1 and
2) is the same, as devices remain active during the same amount of time
(CAP duration).

% \begin{figure}[ht!]
%\begin{center}
% \includegraphics[width=0.8\linewidth]{figures/overhead_steadystate.pdf}
%	\caption{Total control traffic overhead during 6 minutes of network operation in the steady state.}
%	\label{overhead-steadystate}
%\end{center}
%\end{figure}
\vspace{-0.0cm}

\section{Related Work}
\label{relatedwork}

The performance of multi-hop \ieee networks has been well studied during the last years
both using probabilistic approaches \cite{misic} and simulations \cite{anastasi-iscc}. 
Energy consumption introduced during the scan period is widely recognized as a
significant problem. 
The recent work of Karowski \emph{et al.} \cite{karowski}
lowered this cost by optimizing the number of slots to listen over different
channels. 
Romaniello \emph{et al.} \cite{mcbt} proposed the Multichannel Beacon
Train Protocol for faster discovery over multiple channels in the presence
of varying beacon intervals. 
Kohvakka \emph{et al.} discussed a protocol that carries the time offset and the
frequency channel in beacons to ease the scanning process for the joining node
\cite{kohvakka}. It is important to stress that our work in this paper is agnostic of the 
scanning process. Namely, the solicitation scheme we propose starts once a node 
has discovered all neighboring coordinators. 

%In our evaluations, we only use a basic non-optimized scanning procedure on a
%single channel.
%Nevertheless, the proposed scheme of coupling RPL with \textit{beacon-enabled} \ieee is
%compatible with improved scanning techniques proposed in the literature. 
%Thus, the results discussed and derived in Section~\ref{proposal} do not lose
%any generality. 
%In fact, the solicitation scheme we propose starts once a node has discovered
%all neighboring coordinators.

As the de-facto standard for routing in IP-based WSN, RPL has been extensively 
studied in terms of convergence delays, route optimality, path availability, and 
incurred overheads \cite{rpl-loadng, kermejani-iscc}. Coupled with the common
wisdom that cross-layer signaling is necessary for a successful operation of a routing 
protocol in low power and lossy networks, this fact provides a strong support to the 
approach presented in our paper.

The work of Pavkovi\'{c} \emph{et al.} is closely related to ours
\cite{bogdan}. 
The authors proposed the adaptations to the \ieee standard to integrate its
operation with RPL. 
Moreover, they proposed an opportunistic version of RPL to improve the delivery
of time-sensitive traffic and evaluated the proposal in terms of packet delivery
ratio and delay. 
In recent work \cite{bogdan-journal}, they discussed the RPL performance
benefits of modifying the \ieee cluster-tree structure into a ``cluster-DAG''. 
Our work was basically motivated by the same problem---the incompatibility of
two structures, the 802.15.4 cluster-tree and the DODAG.
While the approach of Pavkovi\'{c} \emph{et al.}  presents performance
improvement, its main drawback is the need for modifications of two standards,
RPL and \ieee.
We have addressed the same problem from a different perspective---instead of
modifying the standards, we provide a means for constructing the RPL DODAG and
forming the cluster-tree as its subset. 
As a consequence, we obtain full compliance with both standards.

\vspace{-0.2cm}

\section{Conclusion}
\label{concl}
%Future work to study behavior of RPL downward mechanism (DAOs) over the cluster tree, as well as to add robustness by coupling the RPL DAG with L2 wakeup mechanisms in an efficient way such that node can keep track of possible parents if the preferred parent dies.

We have presented a scheme that allows coupling \textit{beacon-enabled} \ieee with the
RPL routing protocol. 
The scheme does not require any modification to both standards.
We provide a means for RPL to pass the routing information to Layer 2 before the
802.15.4 topology is created by encapsulating RPL DIO messages in beacon frames. 
 The scheme takes advantage of 802.15.4 command frames to solicit DIO messages. 
The effect of the command frames is to reset the Trickle timer that governs
sending of DIO messages. 

We have evaluated the proposed scheme using the Contiki operating system for
constrained nodes and the instruction-level Cooja simulator. 
The results show energy savings during the topology construction phase and in
the steady state.

To isolate evaluation, we have disabled the RPL support for downward traffic
(DAO messages). 
We plan to evaluate its behavior when running on top of the \ieee
cluster-tree as a part of the future work. 
We also plan to consider an optimal L2 mechanism that will allow the
selection of an alternate coordinator from the DODAG in case of link
failures.

% - perspective : Multichannel in BO/SO dynamically adjusted network
% - 
%\end{document} % This is where a 'short' article might terminate
\vspace{-0.3cm}
\section*{Acknowledgments}
%BT We would like to 
%Many thanks to 
%BT all 
%the researchers involved in the development of tools that we used during this
%work, especially Chi-Anh La and Martin Heusse who worked on the Cooja-based  
%emulator.
Many thanks to Jean-Baptiste Guet and Chi-Anh La for the development of tools that we 
used during this work, and to Martin Heusse and anonymous reviewers for comments that 
improved the quality of the paper.
The work of O. Alphand, F. Rousseau, and A. Duda was partially supported by the
French National Research Agency (ANR) project IRIS under contract
ANR-11-INFR-016 and the European Commission FP7 project CALIPSO under contract
288879. The work reflects only the authors views; the European Community is not
liable for any use that may be made of the information contained herein. 

%
% The following two commands are all you need in the
% initial runs of your .tex file to
% produce the bibliography for the citations in your paper.

\bibliographystyle{IEEEtran}
%BT Mettre les auteurs du rapport interne Franck et als.
\balance
\bibliography{rpl-topologyconstruction,biblio_802154} % sigproc.bib is the name of the Bibliography in this case

% Generated by IEEEtran.bst, version: 1.13 (2008/09/30)
\begin{thebibliography}{10}
\providecommand{\url}[1]{#1}
\csname url@samestyle\endcsname
\providecommand{\newblock}{\relax}
\providecommand{\bibinfo}[2]{#2}
\providecommand{\BIBentrySTDinterwordspacing}{\spaceskip=0pt\relax}
\providecommand{\BIBentryALTinterwordstretchfactor}{4}
\providecommand{\BIBentryALTinterwordspacing}{\spaceskip=\fontdimen2\font plus
\BIBentryALTinterwordstretchfactor\fontdimen3\font minus
  \fontdimen4\font\relax}
\providecommand{\BIBforeignlanguage}[2]{{%
\expandafter\ifx\csname l@#1\endcsname\relax
\typeout{** WARNING: IEEEtran.bst: No hyphenation pattern has been}%
\typeout{** loaded for the language `#1'. Using the pattern for}%
\typeout{** the default language instead.}%
\else
\language=\csname l@#1\endcsname
\fi
#2}}
\providecommand{\BIBdecl}{\relax}
\BIBdecl

\bibitem{154}
\emph{{IEEE Std 802.15.4-2011 (Revision of 2006)}}, pp. 1--314, 2011.

\bibitem{oscar}
M.~Vu\v{c}ini\'{c}, B.~Tourancheau, F.~Rousseau, A.~Duda, L.~Damon, and
  R.~Guizzetti, ``{OSCAR: Object Security Architecture for the Internet of
  Things},'' in \emph{WoWMoM}.\hskip 1em plus 0.5em minus 0.4em\relax IEEE,
  2014.

\bibitem{watteyne}
M.~Palattella, N.~Accettura, X.~Vilajosana, T.~Watteyne, L.~Grieco, G.~Boggia,
  and M.~Dohler, ``{Standardized Protocol Stack for the Internet of (Important)
  Things},'' \emph{Communications Surveys Tutorials, IEEE}, vol.~15, no.~3, pp.
  1389--1406, 2013.

\bibitem{rpl-rfc}
T.~Winter, P.~Thubert, A.~Brandt, J.~Hui, R.~Kelsey, P.~Levis, K.~Pister,
  R.~Struik, J.~Vasseur, and R.~Alexander, ``{{RPL}: {IPv6} Routing Protocol
  for Low power and Lossy Networks},'' {IETF}, {RFC} 6550, March 2012.

\bibitem{bogdan-journal}
B.~Pavkovi\'{c}, A.~Duda, W.-J. Hwang, and F.~Theoleyre, ``{Efficient Topology
  Construction for {RPL} over {IEEE} 802.15.4 in Wireless Sensor Networks},''
  \emph{Ad Hoc Networks, Elsevier}, 2013.

\bibitem{trickle-paper}
P.~Levis, N.~Patel, D.~Culler, and S.~Shenker, ``{Trickle: a Self-Regulating
  Algorithm for Code Propagation and Maintenance in Wireless Sensor
  Networks},'' in \emph{NSDI}, vol.~1.\hskip 1em plus 0.5em minus 0.4em\relax
  USENIX Association, 2004, pp. 2--2.

\bibitem{rpl-performance-rfc}
A.~Tripathi, J.~de~Oliveira, and J.~Vasseur, ``{Performance Evaluation of
  Routing Protocol for Low Power and Lossy Networks},'' {IETF}, {RFC} 6687,
  2012.

\bibitem{franck-report}
O.~Alphand, E.~Duble, A.~Duda, M.~Favre, R.~Guizzetti, M.~Heusse, and
  F.~Rousseau, ``{{GreenNet}: a Wireless Sensor Network for Harvested Nodes},''
  Grenoble Informatics Laboratory, Tech. Rep., 2012.

\bibitem{karowski}
N.~Karowski, A.~Viana, and A.~Wolisz, ``{Optimized Asynchronous Multichannel
  Discovery of {IEEE} 802.15.4-Based Wireless Personal Area Networks},''
  \emph{Transactions on Mobile Computing, IEEE}, vol.~12, no.~10, pp.
  1972--1985, 2013.

\bibitem{cooja-paper}
F.~Osterlind, A.~Dunkels, J.~Eriksson, N.~Finne, and T.~Voigt, ``{Cross-Level
  Sensor Network Simulation with {COOJA}},'' in \emph{Local Computer
  Networks}.\hskip 1em plus 0.5em minus 0.4em\relax IEEE, 2006, pp. 641 --648.

\bibitem{kohvakka}
M.~Kohvakka, J.~Suhonen, M.~Kuorilehto, V.~Kaseva, M.~H{\"a}nnik{\"a}inen, and
  T.~D. H{\"a}m{\"a}l{\"a}inen, ``{Energy-Efficient Neighbor Discovery Protocol
  for Mobile Wireless Sensor Networks},'' \emph{Ad Hoc Networks, Elsevier},
  vol.~7, no.~1, pp. 24 -- 41, 2009.

\bibitem{bogdan}
B.~Pavkovi\'{c}, F.~Theoleyre, and A.~Duda, ``{Multipath Opportunistic RPL
  Routing over IEEE 802.15.4},'' in \emph{Proceedings of the MSWiM
  Conference}.\hskip 1em plus 0.5em minus 0.4em\relax ACM, 2011, pp. 179--186.

\bibitem{zigbee-beaconpayload}
Z.~Yiming, Y.~Xianglong, G.~Xishan, Z.~Mingang, and W.~Liren, ``{A Design of
  Greenhouse Monitoring Control System Based on ZigBee Wireless Sensor
  Network},'' in \emph{WiCom}, 2007, pp. 2563--2567.

\bibitem{misic}
J.~Misic, S.~Shafi, and V.~Misic, ``{Performance of a beacon enabled IEEE
  802.15.4 cluster with downlink and uplink traffic},'' \emph{{Parallel and
  Distributed Systems, IEEE Transactions on}}, vol.~17, no.~4, 2006.

\bibitem{anastasi-iscc}
G.~Anastasi, M.~Conti, M.~Di~Francesco, and V.~Neri, ``Reliability and energy
  efficiency in multi-hop ieee 802.15.4/zigbee wireless sensor networks,'' in
  \emph{ISCC}.\hskip 1em plus 0.5em minus 0.4em\relax IEEE, 2010, pp. 336--341.

\bibitem{mcbt}
G.~Romaniello, E.~Potetsianakis, O.~Alphand, R.~Guizzetti, and A.~Duda, ``{Fast
  and Energy-Efficient Topology Construction in Multi-Hop Multi-Channel
  802.15.4 Networks},'' in \emph{WiMob}.\hskip 1em plus 0.5em minus 0.4em\relax
  IEEE, 2013.

\bibitem{rpl-loadng}
M.~Vu\v{c}ini\'{c}, B.~Tourancheau, and A.~Duda, ``{Performance Comparison of
  the RPL and LOADng Routing Protocols in a Home Automation Scenario},'' in
  \emph{WCNC}.\hskip 1em plus 0.5em minus 0.4em\relax IEEE, April 2013, pp.
  1974--1979.

\bibitem{kermejani-iscc}
H.~Kermajani and C.~Gomez, ``Route change latency in low-power and lossy
  wireless networks using rpl and 6lowpan neighbor discovery,'' in
  \emph{ISCC}.\hskip 1em plus 0.5em minus 0.4em\relax IEEE, 2011, pp. 937--942.

\end{thebibliography}
% You must have a proper ".bib" file
% and remember to run:
% latex bibtex latex latex
% to resolve all references
%
% ACM needs 'a single self-contained file'!
%

%\balancecolumns
% That's all folks!
\end{document}